\documentclass[pra]{revtex4}%
\usepackage{amssymb}
\usepackage{graphicx}
\usepackage{amsmath}
\usepackage{amsfonts}%
\providecommand{\U}[1]{\protect\rule{.1in}{.1in}}

\begin{document}
\title{Non-Markovian decay beyond the Fermi Golden Rule: Survival Collapse of the
polarization in spin chains.}
\author{E. Rufeil Fiori}
\email{rufeil@famaf.unc.edu.ar}
\affiliation{\textit{Facultad de Matem\'{a}tica, Astronom\'{\i}a y F\'{\i}sica, Universidad
Nacional de C\'{o}rdoba, Ciudad Universitaria, 5000 C\'{o}rdoba, Argentina.}}
\author{H. M. Pastawski}
\email{horacio@famaf.unc.edu.ar}
\affiliation{\textit{Facultad de Matem\'{a}tica, Astronom\'{\i}a y F\'{\i}sica, Universidad
Nacional de C\'{o}rdoba, Ciudad Universitaria, 5000 C\'{o}rdoba, Argentina.}}

\begin{abstract}
The decay of a local spin excitation in an inhomogeneous spin chain is
evaluated exactly: I) It starts quadratically up to a spreading time $t_{S}.$
II) It follows an exponential behavior governed by a \textit{self-consistent
Fermi Golden Rule}. III) At longer times, the exponential is overrun by an
inverse power law describing return processes governed by quantum diffusion.
At this last transition time $t_{R}$ a \textit{survival collapse} becomes
possible, bringing the polarization down by several orders of magnitude. We
identify this strongly destructive interference as an antiresonance in the
time domain. These general phenomena are suitable for observation through an
NMR experiment.

\end{abstract}
\maketitle

\section{Introduction}

A typical quantum exponential decay \cite{Gam28, GC29} involves a finite set
of states in presence of an \textquotedblleft environment\textquotedblright,
i.e., weakly coupled to a set of states whose spectrum is dense. The decay of
these states is usually described with the Fermi Golden Rule (FGR). However,
this description contains approximations \cite{KF47} that leave aside some
intrinsically quantum behaviors. Various works on models for nuclei, composite
particles \cite{Kha58, FGR78, GMM95}, excited atoms in a free electromagnetic
field \cite{FP99} and in photonic lattices \cite{KKS94}, and models for
decoherence \cite{CPUM98}, showed that the exponential decay has superimposed
beats and does not hold for very short and very long times, compared with the
lifetime of the system. The short time regime has received recent attention in
connection to the Quantum Zeno Effect \cite{CSM77, PU98, FP99, EG00} and has
been observed in trapped atoms \cite{Rai97}. In contrast, although different
models predict some form of\textbf{\ }power law for long times\ \cite{HPZ92,
SV00, DVL05}, the cross over to this long time behavior has been neither
experimentally observed nor physically interpreted.

In this letter, we present a model describing the evolution of a local
excitation in the otherwise homogeneous polarization of a system of
interacting spins. This situation has two desirable properties: 1) The full
dynamics can be solved analytically and interpreted; 2) An actual Nuclear
Magnetic Resonance (NMR) experiment can be tailored to observe this dynamics.
Specifically, our model describes a linear chain of nuclear spins interacting
under an XY (planar) interaction. In this situation, the evolution of a local
spin excitation reduces to the dynamics of a localized density excitation
in\ a system of non-interacting fermions \cite{PLU95, PUL96, DPL04}. This
excitation decays into a well resolved wave packet propagating along the spin
chain. Such decay could be observed with NMR because it also describes the
dynamics of a multiple quantum coherence experiment \cite{BMGP85} in a chain
of spins with dipolar interactions in the solid state \cite{DMF00}.
Furthermore, a full experimental dynamics of an effective XY Hamiltonian is
achieved using NMR pulse sequences in liquid samples where the spin wave
dynamics has been observed \cite{MBS+97}.

Based on our model, we are able to quantify the quantum nature of the
deviations from the Fermi Golden Rule. We identify three well defined time
regimes: \textbf{1)} For short times the decay is quadratic ($1-\left[
tV_{0}/\hbar\right]  ^{2}$), as is expected when the coupling $V_{0}$ of the
local state with the continuum is treated perturbatively. This lasts for a
time $t_{S}\approx\hbar\pi\overline{N}_{1}\left(  \varepsilon_{r}\right)  $,
where $\overline{N}_{1}\left(  \varepsilon_{r}\right)  $ is the spectral
density of the final states at the resonance energy $\varepsilon_{r}$;
\textbf{2)} An intermediate regime characterized by an exponential behavior,
the \textit{self-consistent Fermi Golden Rule} (SC-FGR) where the rate, the
pre-exponential factor and the resonance energy are found self-consistently;
\textbf{3)} A long-time regime in which the exponential law is overrun by an
inverse power law which is identified with the quantum diffusion in the chain.
At this last cross-over, the oscillations could lead to a dip of several
orders of magnitude in the local polarization. This \textit{survival
collapse}\emph{ }is identified with a destructive interference between the
\textit{pure\ survival amplitude}, i.e., the SC-FGR component, and the
\textit{return} amplitude, associated with higher orders in a perturbation
theory. This striking quantum phenomenon can be seen as a dynamical version of
the antiresonance that has been described for steady state observables
\cite{DPW89, Fa61, LPD90}. Now, the destructive interference is also due
to\ the splitting of the wave among two different families of pathways in
space. However, the\ interference is now restricted to the narrow time window
when the amplitudes are comparable and the phases are opposite.

\section{Dynamics in a nuclear spins chain}

We use the known mapping between spins and fermions \cite{LSM61} together with
a new formulation for spin dynamics based on the non-equilibrium Keldysh
formalism \cite{Kel65} developed in Refs. \cite{DPL04} and \cite{DPA05}.

The two spin correlation function in\ a system with $M$ spins\ $1/2$\ evolving
under a Hamiltonian $\hat{H},$%
\begin{equation}
P_{f,i}\left(  t\right)  =\frac{\left\langle \Psi_{\text{eq.}}\left\vert
\hat{S}_{f}^{z}\left(  t\right)  \hat{S}_{i}^{z}\left(  t_{0}\right)
\right\vert \Psi_{\text{eq.}}\right\rangle }{\left\langle \Psi_{\text{eq}%
}\left\vert \hat{S}_{i}^{z}\left(  t_{0}\right)  \hat{S}_{i}^{z}\left(
t_{0}\right)  \right\vert \Psi_{\text{eq}}\right\rangle }, \label{Eq_Pgral}%
\end{equation}
gives the amount of the $z$ component of the local polarization on the site
$f$th at time $t$, provided that the system was, at time $t_{0}\leq t,$ in its
equilibrium state with a spin `up' added at the $i$th site. Here, $\hat{S}%
_{f}^{z}\left(  t\right)  =$e$^{\text{i}\hat{H}t}\hat{S}_{f}^{z}$%
e$^{-\text{i}\hat{H}t}$ is the spin operator in the Heisenberg representation
and $\left\vert \Psi_{\text{eq}}\right\rangle =%
{\textstyle\sum\nolimits_{N}}
s_{N}\left\vert \Psi_{\text{eq}}^{\left(  N\right)  }\right\rangle $ is the
many-body equilibrium mixed state constructed by adding states with different
number $N$ of spins up with the appropriate statistical weights and random
phases. We will assume the high temperature limit, that leads to equal
statistical weights $\left\vert s_{N}\right\vert ^{2}=\frac{1}{2^{M}}\binom
{M}{N}.$

We consider a linear chain of $M$ spins in an external magnetic field. They
interact with their nearest neighbors at distance $a_{o}$ through an
XY\ coupling:%
\begin{equation}
\hat{H}=%
{\displaystyle\sum\limits_{n=0}^{M-1}}
\hbar\Omega_{n}\hat{S}_{n}^{z}-%
{\displaystyle\sum\limits_{n=0}^{M-2}}
\frac{1}{2}J_{n+1,n}\left[  \hat{S}_{n+1}^{x}\hat{S}_{n}^{x}+\hat{S}_{n+1}%
^{y}\hat{S}_{n}^{y}\right]  ,
\end{equation}
where $\hat{S}_{n}^{u}$ $\left(  u=x,y,z\right)  $ represents the Cartesian
spin operator. The first term of this Hamiltonian is the Zeeman energy, where
$\Omega_{n}$ is the chemical shift precession frequency; the second term,
$\hat{H}_{XY}$, contains the $J_{n+1,n}$ coupling between sites $n$ and $n+1.$
It gives the flip-flop interaction $\hat{S}_{n+1}^{+}\hat{S}_{n}^{-}+\hat
{S}_{n+1}^{-}\hat{S}_{n}^{+}$ in terms of the rising and lowering spin
operator $\hat{S}_{n}^{\pm}=\hat{S}_{n}^{x}\pm i\hat{S}_{n}^{y}$.

The Jordan-Wigner (J-W) transformation \cite{LSM61} establishes the relation
between spin and fermion operators at each site $n$. When $\hat{H}$ commutes
with the number operator, the different subspaces $N$ are decoupled. Further
simplification is obtained for Hamiltonians which are quadratic in the
fermionic operator, as the case of $\hat{H}_{XY}$. Due to the short range
interaction, after a J-W transformation, the only non-zero coupling terms are
proportional to $\hat{c}_{n+1}^{+}\hat{c}_{n}=\hat{S}_{n+1}^{+}\hat{S}_{n}%
^{-},$ where $\hat{c}_{n}^{+},$ $\hat{c}_{n}$ are the creation and destruction
operators for fermions. The Hamiltonian become%
\begin{equation}
\hat{H}=%
{\displaystyle\sum\limits_{n=0}^{M-1}}
\varepsilon_{n}\left[  \hat{c}_{n}^{+}\hat{c}_{n}-\frac{1}{2}\right]  -%
{\displaystyle\sum\limits_{n=0}^{M-2}}
V_{n+1,n}\left[  \hat{c}_{n+1}^{+}\hat{c}_{n}+\mathrm{c.c.}\right]  ,
\label{Eq_htb}%
\end{equation}
where $\varepsilon_{n}\equiv\hbar\Omega_{n}$ are the site energies and
$V_{n+1,n}\equiv\frac{1}{2}J_{n+1,n}$ are the hoppings. Each subspace has $N$
non-interacting fermions. The eigenfunctions $\left\vert \Psi_{\gamma
}^{\left(  N\right)  }\right\rangle $\ are expressed as a single Slater
determinant built-up upon the single particle wave functions $\psi_{k}$
describing a particle of energy $\varepsilon_{k}$ in a chain$.$ Under this
condition and defining $\left\vert i\right\rangle \equiv\hat{c}_{i}%
^{+}\left\vert \emptyset\right\rangle ,$ with $\left\vert \emptyset
\right\rangle $ the fermion vacuum, Eq.(\ref{Eq_Pgral}) reduces to
\begin{align}
P_{f,i}\left(  t\right)   &  =\left\vert \left\langle f\right\vert
\exp[-\mathrm{i}\hat{H}t/\hbar]\left\vert i\right\rangle \theta\left(
t\right)  \right\vert ^{2}\label{Eq_Pif}\\
&  \equiv\hbar^{2}\left\vert G_{f,i}^{R}\left(  t\right)  \right\vert ^{2},
\end{align}
where $G_{f,i}^{R}\left(  t\right)  $ is the retarded Green's function for a
single fermion.

Therefore, for systems represented by a 1-d chain of spins with nearest
neighbors XY interaction,\ at high temperature, the dynamics of a
\textit{local polarization amplitude} corresponds exactly to the \textit{wave
function of single particle} evolving according to a tight-binding Hamiltonian.

\section{Local excitation: the exponential decay and beyond}

Let us describe the evolution of a local excitation $\left\vert 0\right\rangle
\equiv\hat{c}_{0}^{+}\left\vert \emptyset\right\rangle $ in a Hamiltonian
whose spectrum has a finite support. This is the case of most excitations in a
lattice. The autocorrelation function is Eq.(\ref{Eq_Pif}) with $\left\vert
i\right\rangle =\left\vert f\right\rangle =\left\vert 0\right\rangle .$
Expanding the initial condition in the eigenstates $\left\vert \psi
_{k}\right\rangle $ one obtains \cite{KF47, Kha58}
\begin{align}
P_{00}(t)  &  =\left\vert \theta\left(  t\right)  \sum_{k=1}^{M}\left\vert
\left\langle \psi_{k}\right\vert \left.  0\right\rangle \right\vert ^{2}%
\exp[-\mathrm{i}\varepsilon_{k}t/\hbar]\right\vert ^{2},\nonumber\\
&  =\left\vert \theta\left(  t\right)  \int_{-\infty}^{\infty}\mathrm{d}%
\varepsilon\text{ }\left[  \sum_{k=1}^{M}\left\vert \left\langle \psi
_{k}\right\vert \left.  0\right\rangle \right\vert ^{2}\text{ }\delta
(\varepsilon-\varepsilon_{k})\right]  \exp[-\mathrm{i}\varepsilon
t/\hbar]\right\vert ^{2}.
\end{align}
The term in brackets is the Local Density of States (LDoS) $N_{0}\left(
\varepsilon\right)  $ at site $0$th. It can be evaluated using the retarded
Green's function,
\begin{align*}
N_{0}\left(  \varepsilon\right)   &  =-\frac{1}{\pi}\operatorname{Im}%
\int\mathrm{d}t\text{ }G_{00}^{R}(t)e^{\mathrm{i}\varepsilon t},\\
&  =-\frac{1}{\pi}\operatorname{Im}G_{00}^{R}(\varepsilon).
\end{align*}
Then, we can\textbf{\ }express the autocorrelation function as the Fourier
transform of the LDoS,
\begin{equation}
P_{00}(t)=\left\vert \theta\left(  t\right)  \int_{-\infty}^{\infty}%
\mathrm{d}\varepsilon\text{ }N_{0}(\varepsilon)\exp[-\mathrm{i}\varepsilon
t/\hbar]\right\vert ^{2}. \label{Eq_Pcomun}%
\end{equation}
This expression has numerical and analytical advantages because the Green's
function can be accurately calculated in the energy representation, and the
integral is limited to the spectral support. Besides, a clear identification
of quantum interferences will be obtained by analyzing the argument under the
modulus operator.

Alternatively, the autocorrelation function can be written as
\begin{equation}
P_{00}(t)=\theta\left(  t\right)  \int_{-\infty}^{\infty}\mathrm{d}%
\omega\text{ }\mathcal{J}_{0}(\omega)\exp[-\mathrm{i}\omega t],
\label{Eq_P(J)}%
\end{equation}
where the spectral density of the particle excitations at site $0$th:%
\begin{equation}
\mathcal{J}_{0}(\omega)=\hbar\int_{-\infty}^{\infty}\mathrm{d}\varepsilon
\text{ }N_{0}(\varepsilon)N_{0}(\varepsilon+\hbar\omega), \label{Eq_J(w)}%
\end{equation}
\ has a direct physical interpretation and can be easily computed \cite{PW87}.

\begin{figure}
[ptb]
\begin{center}
\includegraphics[
height=2.2866in,
width=3.0761in
]%
{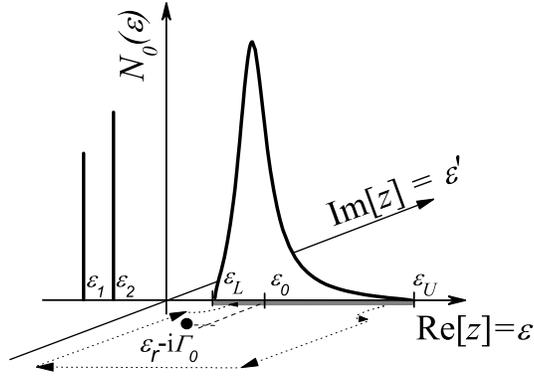}%
\caption{Local spectrum (LDoS) in the complex plane $z=\varepsilon+$i
$\varepsilon\acute{}$. $\varepsilon_{L}$ and $\varepsilon_{U}$ are the lower
and upper band-edges, respectively. $\varepsilon_{1}$ and $\varepsilon_{2}$
are localized states. The resonance energy is $\varepsilon_{r}=\varepsilon
_{0}-\Delta_{0}$, and the pole appears in $\varepsilon_{r}-\mathrm{i}%
\Gamma_{0}.$ The integration path is shown with dotted lines; consist of four
straight lines and two arcs, that avoid the band-edges singularities.}%
\label{fig_camino}%
\end{center}
\end{figure}

All the previous equations remain valid when the size of the system, and hence
the dimension of the Hilbert space, becomes unbounded ($M\rightarrow\infty$).
In this case, either part or the whole of the discrete (pure point) spectrum,
may become a \textit{continuous energy band} of delocalized (extended) states
in the finite range $\left[  \varepsilon_{L},\varepsilon_{U}\right]  $. If the
system does not present localized eigenstates \cite{Ander}, $N_{0}%
(\varepsilon)$ vanishes outside the band (Fig. \ref{fig_camino})\textit{.} On
the other hand, if the initial state $\left\vert 0\right\rangle $ has a finite
weight over one or more localized states its evolution can not fully decay.
Here, we consider cases that exclude such situation. Hence, if $\left\vert
0\right\rangle $ requires an expansion in an infinite number of eigenstates,
its evolution becomes an irreversible decay. In particular, the unperturbed
state of energy $\varepsilon_{0}=\left\langle 0|\hat{H}|0\right\rangle $
becomes a well defined resonance if $\left\vert 0\right\rangle $ can be
expanded\ in terms of the eigenstates within a small breath $\Gamma_{0}$
around an energy $\varepsilon_{r}=\varepsilon_{0}+\Delta_{0}$, where
$\Delta_{0}$ is a small shift due to the interaction. Furthermore, the
validity of the Fermi Golden Rule for $P_{00}(t)$ requires \cite{WVPC02} that
the state $\left\vert 0\right\rangle $ is similarly coupled to each of the
unperturbed states $\left\vert \phi_{k}^{o}\right\rangle $ with energies
$\varepsilon_{k}^{o}$ in a continuum spectrum.%

In order to evaluate the local dynamics, we perform the integral in
Eq.(\ref{Eq_Pcomun}) using the residue theorem and following the path shown in
the Fig. \ref{fig_camino}. In the analytical continuation $N_{0}(z)\equiv
N_{0}(\varepsilon+\mathrm{i}\varepsilon^{\prime})$, resonances appear like
poles in the complex plane. We will consider Hamiltonians where an initially
localized state with energy $\varepsilon_{0}$ interacting with a continuum
gives rise to a single resonance, i.e., the LDoS presents poles at
$\varepsilon_{r}\pm\mathrm{i}\Gamma_{0}$. The van Hove singularities on the
contour are excluded with circle arcs with radii R. Their contribution to the
integral vanish when $R\rightarrow0$, because the band edges are of the form
$\left(  \varepsilon-\varepsilon_{L}\right)  ^{\nu}$ with $\nu>-1.$ Also, the
integral over the contour $z=\varepsilon-\mathrm{i}L;$ $\varepsilon$
$\epsilon$ $[\varepsilon_{L},\varepsilon_{U}],$\ vanish when $L\rightarrow
\infty$. Then, we obtain%
\begin{equation}
P_{00}(t)=|\underset{\mathrm{SC-FGR}}{\underbrace{~a~~\mathrm{e}^{-(\Gamma
_{0}+\mathrm{i}\varepsilon_{r})t/\hbar}}}+\underset{\text{return correction
from quantum diffusion}}{\underbrace{\int\limits_{0}^{\infty}\text{\textrm{d}%
}\varepsilon^{\prime}\mathrm{e}^{-\varepsilon^{\prime}t/\hbar}[\mathrm{e}%
^{-\mathrm{i}\varepsilon_{L}t/\hbar}N_{0}(\varepsilon_{L}-\mathrm{i}%
\varepsilon^{\prime})-\mathrm{e}^{-\mathrm{i}\varepsilon_{U}t/\hbar}%
N_{0}(\varepsilon_{U}-\mathrm{i}\varepsilon^{\prime})]}}|^{2},\label{Eq_Poo}%
\end{equation}
where $a=\lim_{z\rightarrow\varepsilon_{r}-\mathrm{i}\Gamma_{0}}\left[
2\pi\mathrm{i}\text{ }(z-\varepsilon_{r}+\mathrm{i}\Gamma_{0})\text{ }%
N_{0}(z)\right]  $ and $t\geq0$. If we approximate the LDoS by a Lorentzian
function that jumps to zero outside the band, we can see that
\begin{equation}
A\equiv\left\vert a\right\vert ^{2}\simeq1+\delta,
\end{equation}
where
\begin{equation}
0<\delta=\frac{2}{\pi}\frac{\Gamma_{0}}{(\varepsilon_{r}-\varepsilon_{L}%
)}\frac{\left(  \varepsilon_{U}-\varepsilon_{L}\right)  }{(\varepsilon
_{U}-\varepsilon_{r})}\ll1.
\end{equation}
The first term of Eq.(\ref{Eq_Poo}) already supersedes the usual Fermi Golden
Rule approximation since it has a pre-exponential factor ($A\gtrsim1$) and the
exact rate of decay $\Gamma_{0}$. This result is the \textit{self-consistent
Fermi Golden Rule} (SC-FGR). By analogy with a classical Markov chain, this
exponential term is identified with a \textit{\textquotedblleft pure
survival\textquotedblright\ amplitude. }Within the same analogy\textit{,} the
second term will be called\textit{ \textquotedblleft return\textquotedblright%
\ amplitude, }as it is fed upon the initial decay. The first is the dominant
one for a wide range of times, while the diffusive decay of the second,
dominates for long times and brings out the details of the spectral structure
of the system. In the quantum case, the second term is also fundamental for
the normalization at very short times where the most excited energy states of
the whole system can be virtually explored. Both terms combine to provide the
initial quadratic decay (Quantum Zeno regime) required by the perturbation
theory:
\begin{align}
P_{00}\left(  t\right)   &  =1-\frac{t^{2}}{\hbar^{2}}\left\langle
(\varepsilon-\varepsilon_{r})^{2}\right\rangle _{N_{0}}+\cdots,\\
&  =1-\frac{t^{2}}{2!}\left\langle \omega^{2}\right\rangle _{J_{0}}+\cdots.
\end{align}
Here $\left\langle (\varepsilon-\varepsilon_{r})^{2}\right\rangle _{N_{0}}$
and $\left\langle \omega^{2}\right\rangle _{J_{0}}$ are the energy and
frequency second moments of the densities $N_{0}\left(  \varepsilon\right)  $
and $\mathcal{J}_{0}\left(  \omega\right)  $, respectively. This expansion
holds for a time shorter than the spreading time $t_{S}$ of the wave packet
formed by the decay. In other systems, the divergence of the second moment
leads to different short time decays \cite{GRR01}\textbf{.}

For long times, the behavior of $P_{00}\left(  t\right)  $ is governed by the
slowly decaying second term in Eq.(\ref{Eq_Poo}). Only small values of
$\varepsilon^{\prime}$ contribute to the integral. In turn this restricts the
integration of the LDoS to a range near the band-edges. Then, one can go back
to Eq.(\ref{Eq_Pcomun}) and retain only the van Hove singularities of the
Local Density of States at these edges (e.g. $N_{0}(\varepsilon)\varpropto
\theta(\varepsilon-\varepsilon_{L})\left(  \varepsilon-\varepsilon_{L}\right)
^{\nu}$ which implies \cite{PW87}\ $\mathcal{J}_{0}(\omega)\varpropto
\theta(\omega)\omega^{2\nu+1}$). Each singularity would contribute to the slow
decay at long times ($P_{00}(t)\varpropto\left\vert t\right\vert ^{-2\left(
\nu+1\right)  }$). The relative participation of the energy states at each
edge of the LDoS is given by the relative weight of the Lorentzian tails at
these edges:
\begin{equation}
\beta=\frac{(\varepsilon_{r}-\varepsilon_{L})^{2}+\Gamma_{0}^{2}}%
{(\varepsilon_{U}-\varepsilon_{r})^{2}+\Gamma_{0}^{2}}.\label{Eq_beta}%
\end{equation}
Then, the polarization for long times is
\begin{equation}
P_{00}(t)\sim\left[  1+\beta^{2}-2\beta\cos(Bt/\hbar)\right]  \left\vert
\int\text{\textrm{d}}\varepsilon^{\prime}\mathrm{e}^{-\varepsilon^{\prime
}t/\hbar}N_{0}(\varepsilon_{L}-\mathrm{i}\varepsilon^{\prime})\right\vert
^{2},\label{Eq_Plargo}%
\end{equation}
where $B=\varepsilon_{U}-\varepsilon_{L}.$ This means that the long time
behavior is just the power law decay of the integral multiplied by a factor
having a modulation with frequency $B/\hbar.$

\section{Survival collapse}

In steady state transport \cite{DPW89} as well as in dynamical electron
transfer \cite{LPD90} there are situations in which a particle can reach the
final state following two alternative pathways. Since each of them collects a
different phase, this allows a destructive interference blocking the final
state. This phenomenon has been dubbed antiresonance \cite{DPW89, LPD90}. It
extends the Fano resonances, which describes the anomalous ionization
cross-section \cite{Fa61}. In the present case, the survival of the local
excitation also recognizes two alternative pathways: the
\textit{pure\ survival} amplitude which is typically described by the Fermi
Golden Rule, and the paths where the excitation has decayed, explored the
environment, and then returns. These two alternatives can interfere. We
rewrite Eq.(\ref{Eq_Poo}) to emphasize that the local polarization
$P_{00}\left(  t\right)  $ is the result of \textit{two} different
contributions:
\begin{equation}
P_{00}\left(  t\right)  =\left\vert \Psi_{S}+\Psi_{R}\right\vert
^{2}=\left\vert \Psi_{S}\right\vert ^{2}+\left\vert \Psi_{R}\right\vert
^{2}+2\operatorname{Re}[\Psi_{S}^{\ast}\Psi_{R}],
\end{equation}
where the phase in $\Psi_{R}$ arise from the exponentials with $\varepsilon
_{L}$ and $\varepsilon_{U}$ (the LDoS is real for any argument). Hence,
\begin{align}
\Psi_{S}\left(  t\right)   &  =\left\vert a\right\vert \mathrm{e}%
^{-\mathrm{i}\phi_{a}}\mathrm{e}^{-\Gamma_{0}t/\hbar}\mathrm{e}^{-\mathrm{i}%
\left(  \varepsilon_{r}-\varepsilon_{L}\right)  t/\hbar},\\
\Psi_{R}\left(  t\right)   &  =\left\vert \Psi_{R}\left(  t\right)
\right\vert \mathrm{e}^{\mathrm{i}\phi\left(  t\right)  };\\
\phi\left(  t\right)   &  =\arctan\left(  \frac{\beta\sin\left(
Bt/\hbar\right)  }{1-\beta\cos\left(  Bt/\hbar\right)  }\right)
.\label{Eq_phaser}%
\end{align}
where Eq.(\ref{Eq_phaser}) results using the long time limit of
Eq.(\ref{Eq_Plargo}).

While the interference term in $P_{00}\left(  t\right)  $ is present along the
whole exponential regime, it becomes important when both, the \textit{pure
survival}\emph{\ }amplitude and the \textit{return} contribution, are of the
same order. This occurs at the cross-over time $t_{R}$ between the exponential
regime and the power law. The interference term can produce\ a
\textit{survival collapse}, i.e., a pronounced dip that takes $P_{00}\left(
t\right)  $ close to zero (see Fig. \ref{fig_semi}). In order to obtain a full
collapse, two simultaneous conditions are needed;
\begin{align}
\left\vert \Psi_{S}\left(  t_{R}\right)  \right\vert  &  =\left\vert \Psi
_{R}\left(  t_{R}\right)  \right\vert \text{\ and \ }\\
\left(  \varepsilon_{r}-\varepsilon_{L}\right)  t_{R}/\hbar-\phi\left(
t_{R}\right)   &  =\left(  \pi-\phi_{a}\right)  +2\pi n,\text{ }%
n\ \operatorname{integer},\label{Eq_fases}%
\end{align}
which are satisfied with a fair precision because
\begin{equation}
\left\vert \left(  \varepsilon_{r}-\varepsilon_{L}\right)  /\hbar\right\vert
\gg\Gamma_{0}/\hbar>2\pi/t_{R}\geq\left\vert \phi\left(  t_{R}\right)
\right\vert /t_{R},
\end{equation}
i.e., while the return amplitude has a phase with a slow variation, the pure
survival term oscillates rapidly. When both amplitudes are of the same order,
the destructive interference will be noticeable.

\section{Decay in a semi-infinite chain}

Now we focus on\ a specific case of Eq.(\ref{Eq_Pcomun}) that can be achieved
experimentally and has simple analytical properties. We consider the
Hamiltonian of Eq.(\ref{Eq_htb}) with the\ $0$th site (spin) in the chain
different from the others sites in both site energy (chemical shift) and
hopping (J-coupling), i.e., $\varepsilon_{0}\neq\varepsilon_{n}\equiv2V$ and
$V_{0,1}=V_{0}<V_{n,n+1}\equiv V$ for $n>0$. This defines a continuous
spectrum in the range $[0,B\equiv4V]$\textbf{ }which, in the lower edge,
describes a particle of mass $m$ in the continuum with\textbf{ }$V=\hslash
^{2}/(2ma_{o}^{2}).$\ Our model presents a resonance provided that the site
energy is not to close to the band edge, i.e., $\left\vert \varepsilon
_{0}-2V\right\vert <2V-V_{0}^{2}/V$. Otherwise, $\left\vert 0\right\rangle $
would give rise to a localized state \cite{PM01}. The LDoS for this problem is
evaluated using the Dyson equation
\begin{equation}
\left[  G_{0,0}^{R}\left(  \varepsilon\right)  \right]  ^{-1}=\left[
\overline{G}_{0,0}^{R}\left(  \varepsilon\right)  \right]  ^{-1}%
+V_{0,1}\overline{G}_{1,1}^{R}\left(  \varepsilon\right)  V_{1,0},
\end{equation}
following the general continued fraction procedure described in Ref.
\cite{PM01}:
\begin{equation}
N_{0}(\varepsilon)=\frac{1}{2\pi}\frac{\theta\left(  \varepsilon\right)
\theta\left(  4V-\varepsilon\right)  \left(  \tfrac{V_{0}}{V}\right)
^{2}\sqrt{\varepsilon}\sqrt{4V-\varepsilon}}{\left[  \varepsilon
-\varepsilon_{0}-\left(  \tfrac{V_{0}}{V}\right)  ^{2}\left(  \frac
{\varepsilon-2V}{2}\right)  \right]  ^{2}+\left(  \tfrac{V_{0}}{V}\right)
^{4}\left[  V^{2}-\left(  \frac{\varepsilon-2V}{2}\right)  ^{2}\right]
}.\label{Eq_N0}%
\end{equation}
Note that, because of surface effects in the seminfinite $d$-dimensional
space, the LDoS has van Hove singularities of the form $\overline{N}_{1}%
^{(d)}(\varepsilon)\propto\varepsilon^{d/2}$, which differ from those in the
bulk $N_{{}}^{(d)}(\varepsilon)\propto\varepsilon^{(d-2)/2}$. Fig.
\ref{fig_camino} shows $N_{0}(\varepsilon)$ for $V_{0}/V=0.4$ and
$\varepsilon_{0}/V=1.$ The resonant state (the poles of the LDoS) appears in
$\varepsilon_{r}\pm\mathrm{i}\Gamma_{0}$, where
\begin{align}
\varepsilon_{r} &  =\varepsilon_{0}+\Delta_{0};\text{ \ }\Delta_{0}%
=\frac{V_{0}^{2}}{V^{2}-V_{0}^{2}}\frac{\varepsilon_{0}-2V}{2},\\
\Gamma_{0} &  =\frac{V_{0}^{2}}{V^{2}-V_{0}^{2}}\Gamma_{c};\text{ \ }%
\Gamma_{c}=\sqrt{V^{2}-V_{0}^{2}-\left(  \frac{\varepsilon_{0}-2V}{2}\right)
^{2}}.\label{Eq_Gamma0}%
\end{align}
Identifying the local density of states at the first site in absence of
interactions with the $0$-site as
\begin{align}
\overline{N}_{1}(\varepsilon) &  =-\tfrac{1}{\pi}\operatorname{Im}\overline
{G}_{1,1}^{R}(\varepsilon)\nonumber\\
&  =\frac{16}{\pi}\frac{1}{B^{2}}\left(  \frac{1}{2}\sqrt{\varepsilon}%
\sqrt{B-\varepsilon}\right)  \theta\left(  \varepsilon\right)  \theta\left(
B-\varepsilon\right)  \nonumber\\
&  =\frac{16}{\pi}\frac{1}{B^{2}}\Gamma\left(  \varepsilon\right)  .
\end{align}
Note that $a_{o}\Gamma\left(  \varepsilon\right)  /\hbar$ is the group
velocity of a wave packet with energy $\varepsilon$ and $\Gamma\left(
\varepsilon_{0}\right)  \simeq\Gamma_{c}.$ One realizes that the expression of
Eq. (\ref{Eq_N0}) factorizes as a pure Lorentzian and $\overline{N}%
_{1}(\varepsilon):$%
\begin{equation}
N_{0}\left(  \varepsilon\right)  =\frac{V^{2}}{\Gamma_{c}}\frac{\Gamma_{0}%
}{\left(  \varepsilon_{r}-\varepsilon\right)  ^{2}+\Gamma_{0}^{2}}\overline
{N}_{1}(\varepsilon).
\end{equation}
Then, applying the convolution theorem to Eq.(\ref{Eq_Pcomun}) we get a
convolution integral of two functions in the time domain with well
characterized time dependence.
\begin{equation}
G_{00}^{R}\left(  t\right)  =\frac{-\mathrm{i}}{\hbar}\frac{V^{2}}{2\Gamma
_{c}}\theta\left(  t\right)  \int_{-\infty}^{\infty}\mathrm{e}^{-\Gamma
_{0}\left\vert t^{\prime}\right\vert /\hbar}\mathrm{e}^{-\mathrm{i}%
\varepsilon_{r}t^{\prime}/\hbar}g\left(  t-t^{\prime}\right)  \mathrm{d}%
t^{\prime}.
\end{equation}
The first factor inside the integral is the renormalized survival
amplitude\ as described by the SC-FGR. The second factor is the return
amplitude to site $1$ in a semi-infinite chain where site $0$ is missing. It
is expressed in term of the Bessel function of the first kind as $g\left(
t\right)  =2e^{-\mathrm{i}2Vt/\hbar}J_{1}(2Vt/\hbar)/(2Vt/\hbar),$ which shows
fast oscillations and decays with the power law $t^{-3/2}$. This describes the
quantum diffusion in the chain \cite{UP94, DPA05}. It appears convoluted with
an exponential kernel whose oscillation and decay have a longer time scale.
For positive times $g\left(  t\right)  $ coincides with the response function.
This knowledge allows us to solve the integral in the different time regimes
(short, exponential and long time). After some algebra we get
\begin{equation}
P_{00}(t)\approx\left\{
\begin{array}
[c]{c}%
1-\left(  V_{0}t/\hbar\right)  ^{2},\text{ \ \ }t<t_{S}\\
A\exp(-2\Gamma_{0}t/\hbar),\text{ \ \ }t_{S}<t<t_{R}\\
C\left[  1-\frac{2\beta}{1+\beta^{2}}\sin\left(  Bt/\hbar\right)  \right]
\left(  \dfrac{\hbar}{\Gamma(\varepsilon_{r})t}\right)  ^{3},t_{R}<t
\end{array}
\right.  ,\label{Eq_Psemi}%
\end{equation}
where\textbf{\ }$t_{S}$ is the cross-over time from the short time regime to
the exponential SC-FGR, and time $t_{R}$ separates the SC-FGR and the power
law regime. Also,
\begin{align}
A &  =\frac{1}{4\Gamma_{c}^{2}}\sqrt{\varepsilon_{r}^{2}+\Gamma_{0}^{2}}%
\sqrt{\left(  B-\varepsilon_{r}\right)  ^{2}+\Gamma_{0}^{2}}\gtrsim
1,\label{Eq_A}\\
C &  =\frac{\Gamma(\varepsilon_{r})^{3}V}{4\pi\Gamma_{c}^{2}}\frac{\Gamma
_{0}^{2}}{\left(  \Gamma_{0}^{2}+\varepsilon_{r}^{2}\right)  ^{2}}\left[
1+\beta^{2}\right]  \simeq\frac{1}{4\pi}\left(  \frac{V_{0}}{\varepsilon_{0}%
}\right)  ^{2}\left(  \frac{\Gamma_{0}}{\varepsilon_{r}}\right)
,z\label{Eq_c}%
\end{align}
and $\beta$ was defined in Eq.(\ref{Eq_beta}). Here, we used $V_{0}\ll V$ and
$\varepsilon_{0}\simeq\varepsilon_{r}$ $+\ \mathcal{O}(V_{0}^{2}/V)$ in
Eq.(\ref{Eq_Gamma0}) to obtain $\Gamma_{0}\simeq\pi V_{0}^{2}$ $\overline
{N}_{1}\left(  \varepsilon_{r}\right)  $ which coincides with the SC-FGR. Near
the band edge \ $\Gamma_{0}\simeq8V_{0}^{2}/B\sqrt{\varepsilon_{r}/B},$ and
$\Gamma(\varepsilon_{r})\simeq\Gamma_{c}\simeq\sqrt{V\varepsilon_{r}}.$ For
long times, and averaging in a period, one gets
\begin{equation}
P_{00}\left(  t\right)  \simeq\left(  \frac{V_{0}}{\varepsilon_{0}}\right)
^{2}\left(  \frac{1}{4\pi}\frac{\Gamma_{0}}{\varepsilon_{r}}\right)  \left(
\dfrac{\hbar}{\Gamma(\varepsilon_{r})t}\right)  ^{3}.\label{Eq_PooL}%
\end{equation}
At long times, the probability of finding the particle at site $0$ is
proportional to the probability of finding it at site $1$, i.e.,
$P_{00}(t)\simeq\left(  V_{0}/\varepsilon_{0}\right)  ^{2}P_{01}(t).$\emph{
}Hence, the factor gives the probability of tunneling back to $\left\vert
0\right\rangle .$ It meassures how the component of the band edge (that
determine the long time behavior) over the surface state $\left\vert
1\right\rangle $ mixes with state $\left\vert 0\right\rangle $. The
assignation of a time scale to the return probability in the last term is
arbitrary. We choose $\Gamma\left(  \varepsilon_{r}\right)  ,$ the dominant
group velocity of propagating wave packet of energy $\varepsilon_{r}.$ Hence,
the second factor becomes the inverse of the number of cycles within the main
decay.\emph{ }

It is important to note that the cubic power law decay obtained for long times
is a consequence of the $\sqrt{\varepsilon}$ dependence of the band-edges of
the LDoS, i.e., this power law\ is consistent with Eq.(\ref{Eq_Plargo}) taken
together with Eq.(\ref{Eq_N0}). Notice also that the short time scale,
$\hbar/V_{0}$, can also be obtained from the local second moment of the Hamiltonian.

From the analytical result given in Eq.(\ref{Eq_Psemi}), we get the
characteristic times $t_{S}$ and $t_{R}$. A good estimate of $t_{S}$ is
obtained from the minimal distance between the short time decay and the
exponential:
\begin{equation}
\frac{\mathrm{d}}{\mathrm{d}t}\left.  \left[  1-\left(  V_{0}t/\hbar\right)
^{2}-A\exp(-2\Gamma_{0}t/\hbar)\right]  \right\vert _{t=t_{S}}=0.
\end{equation}
Expanding the exponential in its Taylor series, we get
\begin{equation}
t_{S}=\frac{\hbar\Gamma_{0}A}{V_{0}^{2}+2\Gamma_{0}^{2}A}\simeq\hbar
\pi\overline{N}_{1}\left(  \varepsilon_{r}\right)  \simeq\frac{8\hbar}{B}%
\sqrt{\frac{\varepsilon_{r}}{B}},\text{ for }\varepsilon_{r}\ll2V.
\end{equation}
Here, we see that in this parametric regime the short time cross-over is only
determined by $\overline{N}_{1}\left(  \varepsilon_{r}\right)  ,$ the local
density of states at the first site of the chain. We may invoke the optical
theorem \cite{PM01}, to interpret $\overline{N}_{1}\left(  \varepsilon
_{r}\right)  $ as the time scale at which a wave packet with energy
$\varepsilon_{r}$ escapes from the $1$st site region, i.e., the excitation
build from the decay, preventing the return to the original $0$th site
\cite{PM01, thou74}.

The time $t_{R}$ is obtained from the cross-over between the exponential
regime and the power law decay%
\begin{equation}
A\exp(-2\Gamma_{0}t_{R}/\hbar)=C\left(  \dfrac{\hbar}{\Gamma\left(
\varepsilon_{r}\right)  t_{R}}\right)  ^{3}.
\end{equation}
We can use $\sqrt{A/C}\simeq2\sqrt{\pi}\varepsilon_{0}/V_{0}\sqrt
{\varepsilon_{r}/\Gamma_{0}},\ $and solve iteratively the transcendental
equation, i.e.,
\begin{align}
t_{R}^{\left(  0\right)  }  &  =\frac{\hbar}{\Gamma_{0}}\ln\left(  2\sqrt{\pi
}\frac{\varepsilon_{0}}{V_{0}}\sqrt{\frac{\varepsilon_{r}}{\Gamma_{0}}%
}\right)  ,\nonumber\\
t_{R}^{(n+1)}  &  =\frac{\hbar}{\Gamma_{0}}\ln\left(  2\sqrt{\pi}%
\frac{\varepsilon_{0}}{V_{0}}\sqrt{\frac{\varepsilon_{r}}{\Gamma_{0}}}\right)
+\frac{3\hbar}{2\Gamma_{0}}\ln\left[  \frac{\Gamma\left(  \varepsilon
_{r}\right)  }{\hbar}t_{R}^{(n)}\right]  . \label{Eq_tLn1}%
\end{align}
Already in the third order we get a very good agreement with the cross-over
observed in the exact dynamics.

\section{Numerical verification}

\begin{figure}
[ptb]
\begin{center}
\includegraphics[
height=2.4025in,
width=3.3797in
]%
{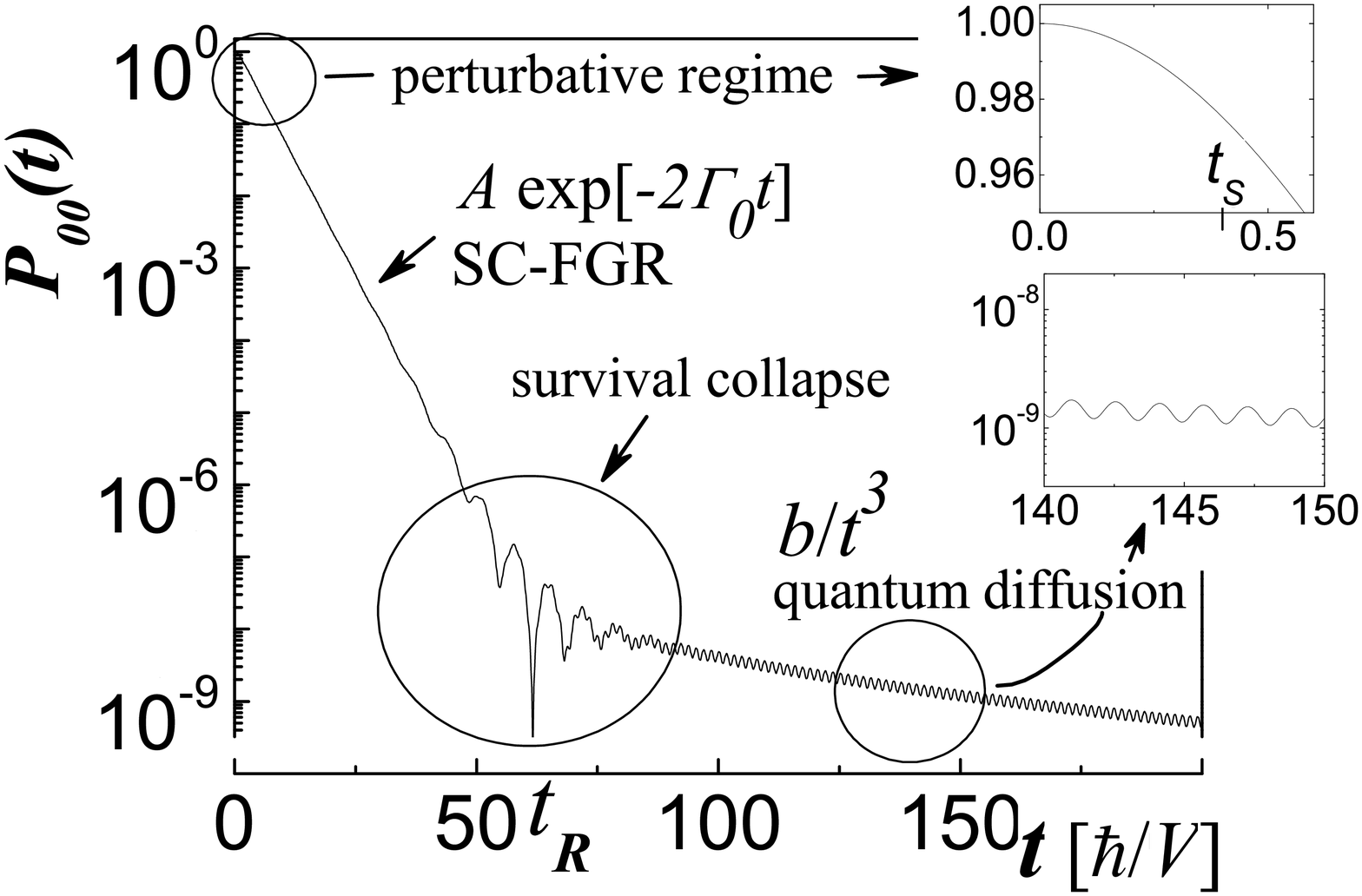}%
\caption{Local polarization, in a semilogarithmic scale, as a function of
time. We consider an unperturbed energy of $\varepsilon_{0}/V=1$ and
interaction strength $V_{0}/V=0.4,$ that leads to a resonance energy
$\varepsilon_{r}/V=0.9$ and an exponential rate $\Gamma_{0}/V=0.14.$ This is
the case that we consider in Fig.\ref{fig_camino}. The decay exhibits: 1) The
quadratic \ perturbative regime, which is shown amplified in the upper inset.
2) The exponential behavior as described by the self-consistent Fermi Golden
Rule. 3) An asymptotic cubic power law decay, where $b=C/\Gamma\left(
\varepsilon_{r}\right)  ^{3}$ (Eq.\ref{Eq_c}). The lower inset \ shows the
oscillation that modulates this decay. The cross-over time $t_{R}$ when the
survival collapse takes place is indicated.}%
\label{fig_semi}%
\end{center}
\end{figure}

We verify the above results following two independent procedures. Since one
has a closed analytical expression for $N_{0}(\varepsilon),$ the numerical
Fourier transform is straightforward. Alternatively, we find the dynamics from
the numerical eigenvalues and eigenvectors\ of the finite system with $M$
sites. Both of them coincide as long as $M$ is big enough so that the finite
system effects become negligible. This requires that the mesoscopic echo
\cite{PUL96}, arising at a time $t_{ME}\approx\hbar M/B,$ appears well beyond
the cross-over time $t_{R}$. Both procedures provide perfect agreement with
the analytical results. In Fig. \ref{fig_semi} we show $P_{00}(t)$ in a
semilogarithmic scale. The exact decay confirms the time dependences exhibited
by the analytical approximation of Eq.(\ref{Eq_Psemi}). The initial quadratic
decay is amplified in the upper inset. Then, the curve is followed by the
exponential SC-FGR. Finally, it presents a cross-over at $t_{R}$ to the
asymptotic power law decay. This time-scale is easily identified through the
survival collapse shown as a dip in the survival probability. There, the
polarization suddenly decreases from its average by almost three orders of
magnitude. The inset on the bottom\ shows the small oscillation that modulates
the power law.%

\begin{figure}
[ptb]
\begin{center}
\includegraphics[
height=2.444in,
width=3.1566in
]%
{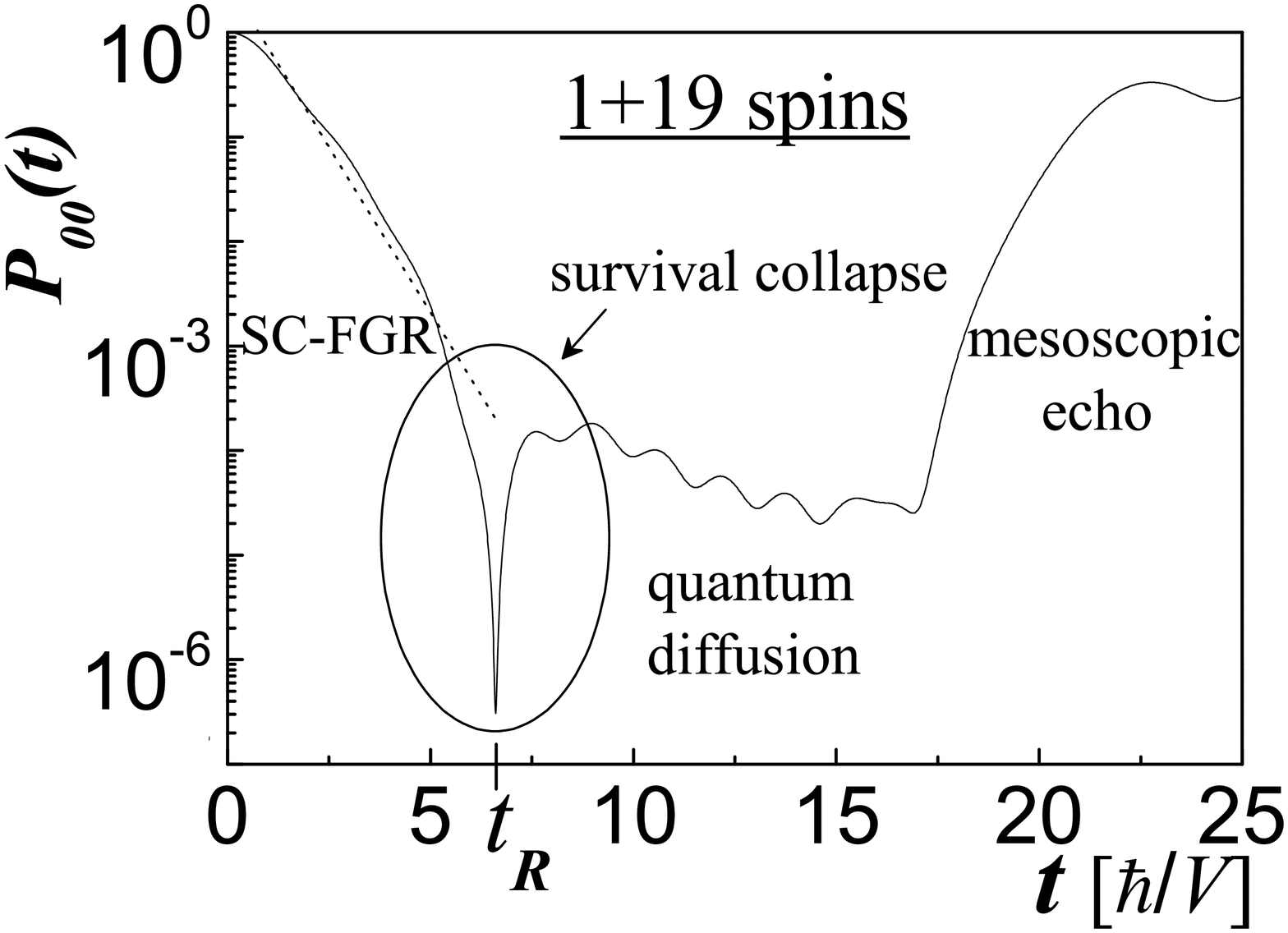}%
\caption{Local polarization, in a semilogarithmic scale, as a function of
time, in units of $\hbar/V$, for $\varepsilon_{0}/V=1.3$, $V_{0}/V=0.75,$ that
leads to a resonance energy $\varepsilon_{r}/V=0.85$ and a SC-FGR exponential,
shown with a dotted line, with rate $\Gamma_{0}/V=0.72$ and pre-exponential
factor $A=2.86$. The \textquotedblleft environment\textquotedblright\ has
$M-1=19$ spins. The decay exhibits a noticeable survival collapse followed by
a cubic power law modulated with a well defined frecuency. At later times, the
mesoscopic echo shows up.}%
\label{Fig_mol}%
\end{center}
\end{figure}

Since the model solved in the previous section could be applied to spins in a
molecule or excitations in a designed nanostructure, both of which have finite
size, it is interesting to verify that the main features discussed also could
be observed in such situations. In Fig. \ref{Fig_mol} we show the dynamics of
one spin in presence of an \textquotedblleft environment\textquotedblright%
\ consisting on a chain of 19 identical 1/2 spins. The three regimes just
discussed are clearly manifested. Later on, it appears a mesoscopic echo at
$tV/\hbar\geq20.$ Note that already at $t_{R}\approx6\hbar/V$ the
magnetization decreases in seven orders of magnitude. For a brief
period\ around $t_{R}$ coherent interference ensure an almost complete
depolarization of the surface site that could not be achieved through
decoherent decay.%

\section{Conclusions}

In the present work we have discussed the exact dynamics of a local excitation
that decays through the interaction with a continuum spectrum with finite
support that acts as an \textquotedblleft environment\textquotedblright. Our
approach goes beyond the usual Markovian approximation that uses the Fermi
Golden Rule to describe these environmental interactions. Within a simple, yet
realistic model of a linear chain of nuclear spins with XY\emph{\ }%
interaction, we found the exact behavior of the autocorrelation function for
all times. The evolution starts with the expected quadratic decay. Then, it
follows the usual exponential FGR regime, but with a corrected rate and a
pre-exponential factor, i.e., the SC-FGR. Finally, we get the long time
regime, that consists of a cubic power law decay modulated by oscillations
whose frequency is determined by the bandwidth. This power law decay is a
consequence of the $\sqrt{\varepsilon}$ behavior of the LDoS in the band-edge
(Eq.(\ref{Eq_Plargo})). A similar result is obtained in models for unstable
nuclei \cite{Kha58, FGR78, GMM95}, an atomic excitation in the free space
\cite{FP99}, and interacting with a photonic band \cite{KKS94}. In those
cases, the decay law is the regular Van Hove singularity of the free space.
Here, the surface modifies the singularity and hence, the time decay. Also, we
found the analytical expressions for the cross-over times $t_{S}$ and $t_{R}$
of Eq.(\ref{Eq_Psemi}) enabling us to assert the range of validity of each regime.

Finally, we find and quantify the survival collapse. This effect, hinted but
not explained in previous works, is visualized as the destructive interference
between the pure survival amplitude and the return amplitude that arises from
pathways that have already explored the rest of the system. This non-Markovian
result fully considers the memory effects to infinite order.

In summary, through the exhaustive solution of a particular model, we made a
conceptual analysis of a general quantum decay process applicable to the great
variety of systems where a quantum exponential decay is observed. Besides this
generality, what gives a particular interest to our model is its suitability
for an experimental test. This would imply the same procedure devised
\cite{MBS+97} to test the mesoscopic echoes \cite{PLU95, PUL96}. In order to
tailor an XY Hamiltonian in an NMR experiment, it uses a radio frequency pulse
sequence\ that produces the truncation of the natural Heisenberg (J-coupling)
Hamiltonian. Alternatively, using the relationship between the dynamics
described by an XY Hamiltonian and multiple quantum dynamics \cite{DMF00}, the
dynamics of our model could be observed with multiple-quantum experiments in
solid state NMR \cite{BMGP85}. The application of one of the above procedures
to relatively small linear molecules would enable the observation of the
survival collapse. One could freeze the dynamics at this time obtaining an
almost null survival of the local excitation. Since the survival collapse
depends critically on the cooperative coherence of the whole system, it would
be quite sensitive to decoherent processes and hence it could be applied to
evaluate them.

\bigskip

\begin{acknowledgments}
\textit{We acknowledge support from Fundaci\'{o}n Antorchas, CONICET, FoNCyT,
and SeCyT-UNC. This work was benefited from discussions with L. Frydman, who
suggested experimental settings where these results are relevant, during a
stay of HMP at the Weizmann Institute of Science. The authors also benefited
from discussions with G. C. Ghirardi and G. Garc\'{\i}a Calder\'{o}n at the
Abdus Salam ICTP and with E. P. Danieli and P.R. Levstein.}
\end{acknowledgments}

\bibliographystyle{elsart-num}
\bibliography{ele_st}

\begin{thebibliography}{10}
\expandafter\ifx\csname url\endcsname\relax
  \def\url#1{\texttt{#1}}\fi
\expandafter\ifx\csname urlprefix\endcsname\relax\def\urlprefix{URL }\fi

\bibitem{Gam28}
G.~Gamow, Z. Phys. 51 (1928) 204--212.

\bibitem{GC29}
R.~W. Gurney, E.~U. Condon, Phys. Rev. 33 (1929) 127--140.

\bibitem{KF47}
N.~S. Krylov, V.~A. Fock, Zh. Eksp. Teor. Fiz. 17 (1947) 93.

\bibitem{Kha58}
L.~A. Khalfin, Sov. Phys. JETP 6 (1958) 1053--1063.

\bibitem{FGR78}
L.~Fonda, G.~C. Ghirardi, A.~Rimini, Rep. Prog. Phys. 41 (1978) 588--630.

\bibitem{GMM95}
G.~Garc{\'{\i}}a-Calder{\'o}n, J.~L. Mateos, M.~Moshinsky, Phys. Rev. Lett. 74
  (1995) 337--340.

\bibitem{FP99}
P.~Facchi, S.~Pascazio, Phys. A 271 (1999) 133--146.

\bibitem{KKS94}
A.~G. Kofman, G.~Kurizki, B.~Sherman, J. Mod. Opt. 41 (1994) 353.

\bibitem{CPUM98}
F.~M. Cucchietti, H.~M. Pastawski, G.~Usaj, E.~Medina, Anales AFA 10 (1998)
  224--227.

\bibitem{CSM77}
C.~B. Chiu, E.~C.~G. Sudarshan, B.~Misra, Phys. Rev. D 16 (1977) 520--529.

\bibitem{PU98}
H.~M. Pastawski, G.~Usaj, Phys. Rev. B 57 (1998) 5017--5020.

\bibitem{EG00}
B.~Elattari, S.~A. Gurvitz, Phys. Rev. Lett. 84 (2000) 2047.

\bibitem{Rai97}
S.~R. Wilkinson, C.~F. Bharucha, M.~C. Fischer, K.~W. Madison, P.~R. Morrow,
  Q.~Niu, B.~Sundaram, M.~G. Raizen, Nature 387 (1997) 575--577.

\bibitem{HPZ92}
B.~L. Hu, J.~P. Paz, Y.~Zhang, Phys. Rev. D 45 (1992) 2843.

\bibitem{SV00}
J.~Stolze, M.~Vogel, Phys. Rev. B 61 (2000) 4026--4032.

\bibitem{DVL05}
D.~P. DiVicenzo, D.~Loss, Phys. Rev. B 71 (2005) 035318.

\bibitem{PLU95}
H.~M. Pastawski, P.~R. Levstein, G.~Usaj, Phys. Rev. Lett. 75 (1995)
  4310--4313.

\bibitem{PUL96}
H.~M. Pastawski, G.~Usaj, P.~R. Levstein, Chem. Phys. Lett. 261 (1996)
  329--334.

\bibitem{DPL04}
E.~P. Danieli, H.~M. Pastawski, P.~R. Levstein, Chem. Phys. Lett. 384 (2004)
  306--311.

\bibitem{BMGP85}
J.~Baum, M.~Munowitz, A.~N. Garroway, A.~Pines, J. Chem. Phys. 83 (1985)
  2015--2026.

\bibitem{DMF00}
S.~I. Doronin, I.~I. Maksimov, E.~B. Fel'dman, Zh. Eksp. Teor. Fiz.

\bibitem{MBS+97}
Z.~L. M{\'a}di, B.~Brutsher, T.~Schulte-Herbr{\"u}ggen, R.~Br{\"u}schweiler,
  R.~R. Ernst, Chem. Phys. Lett. 268 (1997) 300--305.

\bibitem{DPW89}
J.~L. D'Amato, H.~M. Pastawski, J.~F. Weisz, Phys. Rev. B 39 (1989) 3554--3562.

\bibitem{Fa61}
U.~Fano, Phys. Rev. 124 (1961) 1866.

\bibitem{LPD90}
P.~R. Levstein, H.~M. Pastawski, J.~L. D'Amato, J. Phys. Cond. Matt. 2 (1990)
  1781--1794.

\bibitem{LSM61}
E.~H. Lieb, T.~Schultz, D.~C. Mattis, Ann. Phys. (N.Y.) 16 (1961) 407.

\bibitem{Kel65}
L.~V. Keldysh, Zh. Eksp. Teor. Fiz. 47 (1964) 1515 [Sov. Phys. JEPT 20 (1965)
  335].

\bibitem{DPA05}
E.~P. Danieli, H.~M. Pastawski, G.~A. \'{A}lvarez, Chem. Phys. Lett. 402 (2005)
  88--95.

\bibitem{PW87}
H.~M. Pastawski, C.~Wiecko, Phys. Rev. A 36 (1987) 5854.

\bibitem{Ander}
P.~W. Anderson, Rev. of Mod. Phys. 50 (1978) 191, section II.

\bibitem{WVPC02}
D.~A. Wisniacki, E.~G. Vergini, H.~M. Pastawski, F.~M. Cucchietti, Phys. Rev. E
  65 (2002) 055206, Eq. 4.

\bibitem{GRR01}
G.~Garc{\'{\i}}a-Calder{\'o}n, V.~Riquer, R.~Romo, J. Phys. A 34 (2001)
  4155--4165.

\bibitem{PM01}
H.~M. Pastawski, E.~Medina, Rev. Mex. de F\'{\i}s. 47 (2001) 1--23.

\bibitem{UP94}
G.~Usaj, H.~M. Pastawski, Anales de la AFA 6 (1994) 155--157.

\bibitem{thou74}
D.~J. Thouless, Phys. Rep. 13 (1974) 93.

\end{thebibliography}

\bigskip
\end{document}